\documentclass[reprint,superscriptaddress,showpacs,amsmath,amssymb,aps,prl]{revtex4-1}

\usepackage{graphicx}
\usepackage{dcolumn}
\usepackage{bm}
\usepackage[colorlinks=True,urlcolor=blue,linkcolor=blue,citecolor=blue]{hyperref}
\usepackage{xcolor}
\mathchardef\mhyphen="2D

\begin{document}


\title{Bulk Signatures of Pressure-Induced Band Inversion and Topological Phase Transitions in Pb$_{1-x}$Sn$_x$Se}

\author{Xiaoxiang Xi}
\affiliation{Photon Sciences, Brookhaven National Laboratory, Upton, New York 11973, USA}

\author{Xu-Gang He}
\affiliation{Condensed Matter Physics and Materials Science Department, Brookhaven National Laboratory, Upton, New York 11973, USA}
\affiliation{Department of Physics and Astronomy, Stony Brook University, Stony Brook, New York 11794, USA}

\author{Fen Guan}
\affiliation{Department of Physics and Astronomy, Stony Brook University, Stony Brook, New York 11794, USA}

\author{Zhenxian Liu}
\affiliation{Geophysical Laboratory, Carnegie Institution of Washington, Washington D.C. 20015, USA}

\author{R. D. Zhong}
\affiliation{Condensed Matter Physics and Materials Science Department, Brookhaven National Laboratory, Upton, New York 11973, USA}

\author{J. A. Schneeloch}
\affiliation{Condensed Matter Physics and Materials Science Department, Brookhaven National Laboratory, Upton, New York 11973, USA}

\author{T. S. Liu}
\affiliation{Condensed Matter Physics and Materials Science Department, Brookhaven National Laboratory, Upton, New York 11973, USA}
\affiliation{School of Chemical Engineering and Environment, North University of China, Taiyuan 030051, China}

\author{G. D. Gu}
\affiliation{Condensed Matter Physics and Materials Science Department, Brookhaven National Laboratory, Upton, New York 11973, USA}

\author{D. Xu}
\affiliation{Department of Physics and Astronomy, Stony Brook University, Stony Brook, New York 11794, USA}

\author{Z. Chen}
\affiliation{Department of Geosciences, Stony Brook University, Stony Brook, New York 11794, USA}

\author{X. G. Hong}
\affiliation{Department of Geosciences, Stony Brook University, Stony Brook, New York 11794, USA}

\author{Wei Ku}
\affiliation{Condensed Matter Physics and Materials Science Department, Brookhaven National Laboratory, Upton, New York 11973, USA}

\author{G. L. Carr}
\affiliation{Photon Sciences, Brookhaven National Laboratory, Upton, New York 11973, USA}
\date{\today}

\begin{abstract}
The characteristics of topological insulators are manifested in both their surface and bulk properties, but the latter remain to be explored. Here we report bulk signatures of pressure-induced band inversion and topological phase transitions in Pb$_{1-x}$Sn$_x$Se ($x=0.00$, 0.15, and 0.23). The results of infrared measurements as a function of pressure indicate the closing and the reopening of the band gap as well as a maximum in the free carrier spectral weight. The enhanced density of states near the band gap in the topological phase give rise to a steep interband absorption edge. The change of density of states also yields a maximum in the pressure dependence of the Fermi level. Thus our conclusive results provide a consistent picture of pressure-induced topological phase transitions and highlight the bulk origin of the novel properties in topological insulators.

\end{abstract}
\maketitle

Topological insulators (TIs) represent a new phase of matter that completely defies Landau's spontaneous symmetry breaking framework \cite{Hasan2010,Qi2011}. Their insulating bulk is topologically nonequivalent to the vacuum, giving rise to metallic surface states. Such exotic properties come with a conundrum: The topological invariants characterizing the topological property are defined using the bulk electronic wavefunctions, but nevertheless remain elusive to bulk experimental probes. To date, the proof of the existence of TIs is based on the detection of the metallic surface states \cite{Hasan2010,Qi2011}. Exploiting the surface spin texture, spin-resolved angle-resolved photoemission spectroscopy provides the only measure of the topological invariants in 3D TIs \cite{Hsieh2009,Xu2011}. Despite the great amount of research activity in this field, the bulk properties of TIs that are intimately tied to the nontrivial topology are mostly unexplored. Establishing the bulk properties can complete our understanding of TIs and facilitate their identification.

An effective approach for establishing the bulk signatures of TIs is to follow the evolution of characteristic features, starting from the trivial insulating state through the topological phase transition
(TPT) and into the TI phase. Applying pressure offers a particularly attractive method for controllably driving a material through such a transition. Generally, a hallmark of a TPT in a non-interacting system is band inversion \cite{Zhang2009}: the bulk band gap closes at the phase transition and reopens afterwards, inverting the characters of the bottom conduction band and top valence band. Although the resulting change in the bulk band structure is expected to be dramatic, detecting and understanding the associated experimental signatures are surprisingly challenging. For example, angle-resolved photoemission spectroscopy is not compatible with pressure tuning nor is it sensitive to the bulk. Previously, two groups \cite{Xi2013,Tran2014} reported investigations for a pressure-induced TPT in BiTeI \cite{Bahramy2012}, but reached different and actually contradictory conclusions. In one case \cite{Xi2013}, an observed maximum in the free carrier spectral weight was interpreted as strong evidence for a TPT.  In the other \cite{Tran2014}, a monotonic redshift of the interband absorption edge was interpreted to indicate the absence of such a transition. Obviously, resolving this contradiction is necessary for our understanding of TIs to move forward.   

Here we clarify this current controversy by demonstrating bulk signatures of a pressure-induced band inversion and thus a TPT in Pb$_{1-x}$Sn$_x$Se ($x=0.00$, 0.15, and 0.23). A maximum in the free carrier spectral weight is reconfirmed in this system and is possibly a generic feature of pressure-induced TPTs when bulk free carriers are present. The absorption edge initially redshifts and then blueshifts under pressure, but only when its overlap with the intraband transition is suppressed. Extra evidence for the TPT is uncovered, including a steeper absorption edge in the topological phase compared to the trivial phase and a maximum in the pressure dependence of the Fermi level. The TPTs in Pb$_{1-x}$Sn$_x$Se imply the creation of 3D Dirac semimetals at the critical pressure, serving as a route for pursuing Weyl semimetals \cite{Gorbar2013}. The robust bulk signatures of TPTs identified here are expected to be useful for exploring a variety of candidate pressure-induced TIs \cite{Chadov2010,Sa2011,Li2014} and even topological materials with electronic correlation \cite{Yang2010,Kim2012}. 

Lead chalcogenides are candidate topological crystalline insulators (TCIs) under pressure \cite{Barone2013}, with the role of time-reversal symmetry in TIs replaced by crystal symmetries \cite{Fu2011}. Similar to TIs, TCIs' nontrivial band topology is associated with an inverted band structure, shown in SnTe \cite{Hsieh2012,Tanaka2012}, Pb$_{0.4}$Sn$_{0.6}$Te \cite{Xu2012}, and Pb$_{0.77}$Sn$_{0.23}$Se below 100 K \cite{Dziawa2012}. These narrow-gap semiconductors crystallize in the rocksalt structure and share simple band structures ideal for investigating bulk characteristic of TPTs. At ambient condition, Pb$X$ ($X$ = S, Se, or Te) has a direct band gap at the L point of the Brillouin zone, with the L$_6^-$ (L$_6^+$) character for the bottom conduction band (top valence band) \cite{Barone2013}. Band inversion is known to be induced in Pb$_{1-x}$Sn$_xX$ ($X$ = Se and Te) by doping \cite{Dimmock1966,Strauss1967,Harman1969} or, in the series of Pb-rich alloys, by cooling \cite{Strauss1967,Harman1969}, although the connection to band topology was only recently recognized \cite{Hsieh2012,Barone2013}. Pressure-induced band inversion in PbSe and PbTe has been proposed on theoretical grounds \cite{Svane2010} especially in the context of TPTs \cite{Hsieh2012,Barone2013}, but experimentally it has not been firmly established \cite{Martinez1973,Akimov1978}.

In this work we present a systematic infrared study of pressure-induced band inversion in Pb$_{1-x}$Sn$_x$Se. Samples with nominal $x=0.00$, 0.15, and 0.23 were synthesized by a modified floating zone method. Hall effect measurements determined a hole density of roughly 10$^{18}$~cm$^{-3}$ for PbSe at room temperature. High-pressure experiments were performed using diamond anvil cells at Beamline U2A of National Synchrotron Light Source, Brookhaven National Laboratory. Samples in the form of thin flakes ($\le 5~\mu$m) were measured in transmission while thicker ($> 10~\mu$m) pieces were measured in reflection. Details of the measurement procedures are described elsewhere \cite{Xi2013}. 

For comparison with the experiment, an \textit{ab initio} method based on the WIEN2k package \cite{wien2k} was used to simulate the pressure effects on PbSe. We combined the local spin density approximation and spin-orbit coupling for the self-consistent field calculations, adjusting only the lattice parameter $a$ to mimic pressure effects. The mesh was set to 46$\times$46$\times$46 k-points and R$_{\mathrm{MT}}\times$K$_{\mathrm{MAX}}=7$, where R$_{\mathrm{MT}}$ is the smallest muffin tin radius and K$_{\mathrm{MAX}}$ the plane wave cutoff. The effect of doping was considered in the self-consistent calculation according to the experimental hole density. While the calculation does not yield the exact lattice parameter at which the TPT occurs, we found the results to be in qualitative agreement with our experimental observations.

\begin{figure}[t]
\includegraphics[scale=0.9]{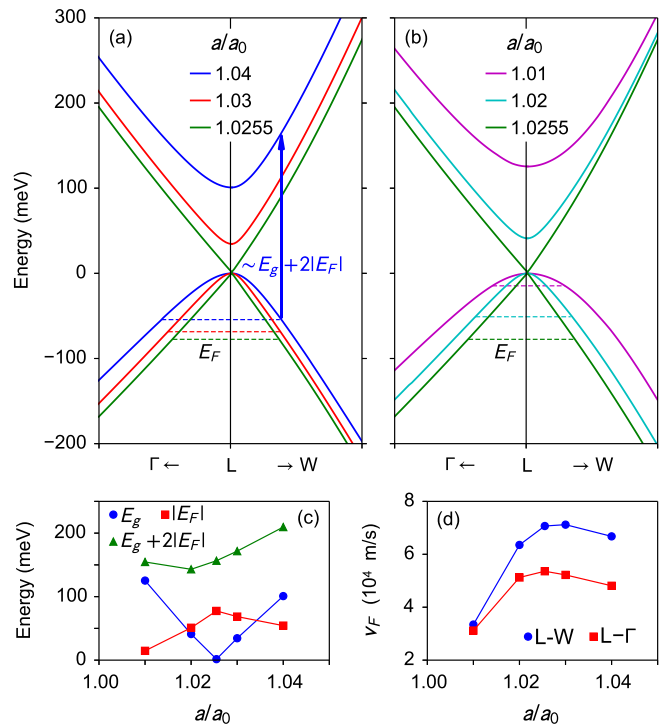}   
\caption{(color online). (a,b) Electronic band structure of PbSe at various lattice parameter ratio $a/a_0$ along the L-$\Gamma$ and L-W directions of the Brillouin zone. $a_0$ is the experimental lattice constant in the zero temperature limit and at ambient pressure \cite{Preier1979}. The TPT occurs at $a/a_0\approx 1.0255$. The dashed lines indicate the Fermi level $E_F$ for a hole density of $N=10^{18}$ cm$^{-3}$. (c) The direct band gap $E_g$ at the L point, absolute value of the Fermi level $|E_F|$ relative to the top valence band, and $E_g+2|E_F|$ [roughly the energy threshold for direct interband transitions in the presence of free carriers, as indicated by the arrow in (a)] as a function of $a/a_0$. (d) Fermi velocity $v_F$ along the L-$\Gamma$ and L-W directions as a function of $a/a_0$ for $N=10^{18}$ cm$^{-3}$.} 
\label{FIG1}
\end{figure}

\begin{figure*}[t]
\includegraphics[scale=0.98]{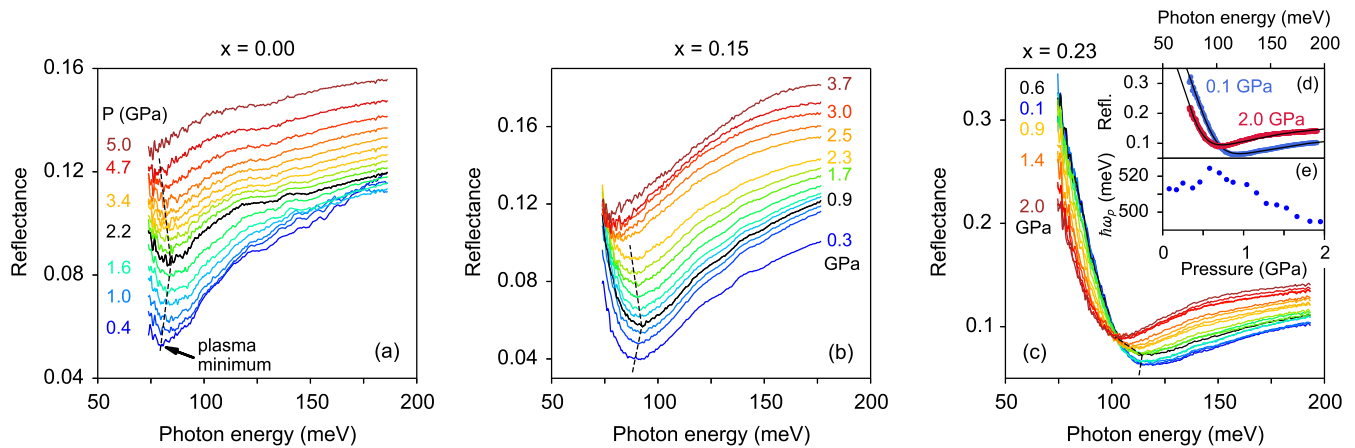}   
\caption{(color online). (a--c) Pressure-dependent mid-infrared reflectance of Pb$_{1-x}$Sn$_x$Se measured at the diamond-sample interface and at room temperature \cite{fringe}. The dashed lines are guides to the eye for the shift of the plasma minimum. (d) Example fits (solid lines) to the experimental data of $x=0.23$ (dots). (e) Pressure dependence of $\omega_p$ extracted from the fit.} 
\label{FIG2}
\end{figure*}

Lead chalcogenides exhibit structural phase transitions under pressure \cite{Ovsyannikov2007}. We determined that, at room temperature, Pb$_{1-x}$Sn$_x$Se maintains the ambient-pressure structure up to 5.1, 4.0, and 2.9 GPa for $x=0.00$, 0.15, and 0.23, respectively \cite{SuppMat}. In the following we focus on the ambient-pressure structure and investigate pressure-induced TPTs. 

We began by revealing a maximum in the pressure-dependence of the bulk free carrier spectral weight. In the same manner as for BiTeI \cite{Xi2013}, this serves as a signature of band inversion. Though the free carrier response of the bulk has made analysis of the surface behavior challenging, it actually provides a sensitive and convenient probe of band inversion and thus TPTs. For doped semiconductors with simple bands, the low-frequency dielectric function for intraband transitions can be described by $\varepsilon(\omega) = \varepsilon_{\infty}-\omega_p^2/(\omega^2+i\omega\gamma)$, where $\omega$ is the photon frequency, $\varepsilon_{\infty}$ the high-energy dielectric constant, $\omega_p^2/8$ the Drude spectral weight ($\omega_p$ the bare plasma frequency), and $\gamma$ the electronic scattering rate. The Drude weight $\omega_p^2/8$ connects with the band dispersion through the Fermi velocity $v_F$, $\omega_p \propto v_F$, because $\omega_p^2$ (in general as a tensor) for a single band is given by \cite{SuppMat}
\begin{equation}
\omega_{p,\alpha\beta}^2=\frac{{\hbar}^2e^2}{\pi}\int d\mathbf{k}\:v_{\alpha}(\mathbf{k})v_{\beta}(\mathbf{k})\,\delta\left(E(\mathbf{k})-E_F\right),
\nonumber
\end{equation}
where $\mathbf{k}$ is the crystal momentum, $E(\mathbf{k})$ the band dispersion, $v_{\alpha}({\mathbf{k}})=[\partial E(\mathbf{k})/\partial k_{\alpha}]/\hbar$ the $\alpha$th component of the Bloch electron mean velocity, and $E_F$ the Fermi energy. The measured reflectance has a minimum near the zero crossing in the real part of $\varepsilon(\omega)$, called the plasma minimum, located at $\omega\sim\omega_p/\sqrt{\varepsilon_{\infty}}$ \cite{Yu2010}. The plasma minimum is universally observed in Pb$_{1-x}$Sn$_{x}$Se for $x=0.00$, 0.15, and 0.23 (see Fig.~\ref{FIG2}). Upon increasing pressure, it initially blueshifts and then redshifts, indicated by the dashed lines in Fig.~\ref{FIG2}. Since the phase space for intraband transitions reaches a minimum at the gap-closing pressure and $\varepsilon_{\infty}$ increases monotonically under pressure \cite{SuppMat}, the maximum in $\omega_p/\sqrt{\varepsilon_{\infty}}$ must be attributed to $v_F$ going through a maximum near the critical pressure of the TPT [Fig.\ref{FIG1}(d)]. The expression for $\varepsilon(\omega)$ provides excellent fits to our data, exemplified for Pb$_{0.77}$Sn$_{0.23}$Se in Fig.~\ref{FIG2}(d) and quantifying the maximum in the pressure dependence of $\omega_p$ [Fig.~\ref{FIG2}(e) and Ref. \cite{SuppMat}]. Such a maximum was also observed in BiTeI \cite{Xi2013,Tran2014} and is likely generic in pressure-induced TPTs when a significant carrier density exists. 

\begin{figure*}[t]
\includegraphics[scale=0.98]{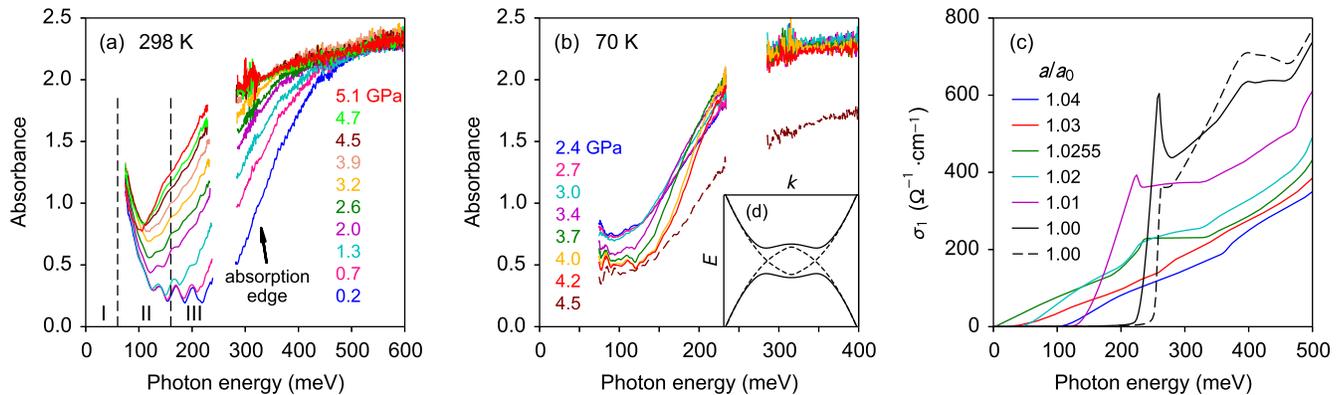}   
\caption{(color online). Pressure-dependent mid-infrared absorbance of PbSe measured at (a) 298 K and (b) 70 K \cite{fringe}. Data in the blank region between 200--300 meV are not shown because of unreliability caused by diamond absorption. (c) Real part of the interband optical conductivity $\sigma_1$ of intrinsic PbSe at various $a/a_0$ from first-principles calculations. The dashed line shows the result including holes with a density of 10$^{18}$ cm$^{-3}$. (d)[inset to (b)] A diagram illustrating hybridization opens up a band gap (in the bands shown as solid lines) when the conduction band and valence band cross (dashed lines).} 
\label{FIG3}
\end{figure*}

Having established the TPTs, we now turn to the interband transitions to address the controversy over the absorption edge \cite{Xi2013,Tran2014}. Fig.~\ref{FIG3}(a) shows the absorbance [defined as $-$log(transmittance)] of PbSe at 298 K for pressures up to $\sim$5.1 GPa, at which point a structural phase transition occurs \cite{Note}. To assist the discussion, we roughly define three photon energy regions, illustrated in Fig.~\ref{FIG3}(a). Region I is dominated by the intraband transition, but is outside the spectral range for the instrument used in the measurement. Region III hosts the majority of the absorption edge, defined as the steep rising part due to the onset of interband transitions and indicated by the arrow in Fig.~\ref{FIG3}(a), which is expected to redshift and then blueshift across the TPT. The absorption edge shown in Region III of Fig.~\ref{FIG3}(a) redshifts monotonically under pressure, indicating band gap closing, but not reopening. Above 1.3 GPa, the initial rising part of the absorption edge which determines the band gap moves into Regions II and I, overlapping significantly with the intraband transition.

Close inspection of Region II in Fig.~\ref{FIG3}(a) reveals band gap reopening. Despite of the overlap with the conspicuous tail of the intraband transition, the interband absorption edge in Region II shows a clear change of slope: it becomes steeper as the pressure is increased to 5.1 GPa, possibly due to the band gap reopening. For a more conclusive observation of the band gap reopening, we cooled the sample to 70 K in order to reduce the electronic scattering rate $\gamma$, so that the intraband transition peak became narrower and overlapped less with the interband absorption edge. As shown in Fig.~\ref{FIG3}(b), the absorption edge systematically tilts towards higher photon energy from 2.4 to 4.2 GPa, suggesting that the band gap monotonically increases. Finally, a structural phase transition occurs at 4.5 GPa (at 70 K), causing a dramatic overall decrease of absorbance. Pressure-induced band gap closing and reopening were also observed in PbTe at low temperature \cite{SuppMat}. 

The above discussion illustrates the complexity of analyzing the interband absorption edge to identify gap closure and band inversion at a TPT.  Considering the apparent monotonic increase of spectral weight in Region III [see Fig.~\ref{FIG3}(a)] as a function of pressure, one might conclude that a TPT had not occurred. But the key signature of gap closure is obscured by overlap with the intraband absorption, as well as by thermal broadening. This can be circumvented by cooling the material, revealing both the redshift and then blueshift of the absorption edge as the band gap closes and then reopens. The situation for BiTeI is even more complicated due to the Rashba splitting and the additional optical transitions among the split subbands \cite{Xi2013,Tran2014}. Cooling does not alleviate this complication.  Thus, inferring how the band gap changes in that and similar materials from measurements to sense the absorption edge is not practical.

In the rest of this Letter, we present two more signatures of pressure-induced TPTs in PbSe, namely a steeper absorption edge in the topological phase and a maximum in the pressure dependence of the Fermi level.

The absorption edge becomes steeper after the TPT, distinguishing the topological phase from the trivial phase. Such behavior is clearly observed in Fig.~\ref{FIG3}(b) and confirmed by the calculated optical conductivity shown in Fig.~\ref{FIG3}(c). The results emphasize the hybridization nature of the band gap in a TI (or TCI) as illustrated in Fig.~\ref{FIG3}(d), qualitatively different from that in a trivial insulator. As demonstrated by the evolution of the band structure across a TPT close to the direct band gap, shown in Fig.~\ref{FIG1}(a--b), before the TPT, pressure suppresses the band gap and transforms the band dispersion from a near-parabolic shape to almost linear. After the TPT, the band dispersion briefly recovers the near-parabolic shape and then flattens. [At even higher pressure, it develops a Mexican-hat feature \cite{SuppMat} similar to that illustrated in Fig.~\ref{FIG3}(d).] The flat band makes the joint density of states just above the band gap much greater than that of an ordinary insulator with the same band gap size, yielding a steeper absorption edge. It also gives rise to Van Hove singularities that differ from the typical ones \cite{He2014}, shown as peaks in the optical conductivity for $a/a_0=1.01$ and 1.00 in Fig.~\ref{FIG3}(c). 

Such a peak feature was previously observed (although unexplained) in the Bi$_2$Te$_2$Se TI material with low free carrier density \cite{Pietro2012,Akrap2012}, but is absent in our infrared absorbance data shown in Fig.~\ref{FIG3}(a--b), possibly for two reasons. First, the peak only appears deep in the topological phase, which requires a high pressure that in reality causes a structural phase transition. Second, the Burstein-Moss effect (see the next paragraph) in our sample precludes optical transitions connecting the states near the top valence band and the bottom conduction band and thus the observation of Van Hove singularities. The dashed line in Fig.~\ref{FIG3}(c) demonstrates that holes with a density of 10$^{18}$~cm$^{-3}$ completely smears the sharp peak. 

Lastly, our calculation shows a maximum in the pressure-dependence of the Fermi level $|E_F|$ at the gap-closing pressure [Fig.~\ref{FIG1}(c)], which can be measured from Shubnikov-de Haas oscillations \cite{VanGennep2014} to support the TPT. This maximum in $|E_F|$ happens because the density of states near the top valence band diminishes as the band dispersion becomes linear \cite{SuppMat}, pushing the Fermi level away from the top valence band to conserve the phase space for the holes. This effect also manifests in the infrared spectra, although $E_F$ cannot be easily determined from them. When free carriers are present, the band gap associated with the absorption edge is not the true band gap in the electronic band structure. As illustrated in Fig.~\ref{FIG1}(a), the holes shift the Fermi level to below the top valence band, making the energy threshold for direct interband transitions approximately $E_g+2|E_F|$, known as the Burstein-Moss effect. The absorption edge characterizes $E_g+2|E_F|$ instead of $E_g$. The combined pressure effects on $E_g$ and $E_F$ retain a minimum in $E_g+2|E_F|$, but the corresponding pressure could be different from the critical pressure for band gap closing, shown in Fig.~\ref{FIG1}(c). Moreover, the absorption edge never redshifts to zero photon energy even when $E_g=0$. The Burstein-Moss effect adds further difficulty to the identification of TPTs using the absorption edge.  

To summarize, we have established bulk signatures of pressure-induced band inversion and thus topological phase transitions in Pb$_{1-x}$Sn$_x$Se ($x=0.00$, 0.15, and 0.23). Infrared reflectance shows a maximum in the bulk free carrier spectral weight near the gap-closing pressure. The interband absorption edge tracks the change of the band gap across the topological phase transition, however the free carriers complicate the picture due to the overlap with the intraband transition and the shift of the Fermi level. The absorption edge becomes steeper in the topological phase due to the hybridization nature of the band gap in topological insulators. A maximum in the pressure dependence of the Fermi level is also expected. These robust bulk features complement the surface-sensitive techniques and serve as a starting point to investigate topological phase transitions in more complicated systems. 

We are grateful for helpful discussions with C. C. Homes. This work was supported by the U. S. Department of Energy through contract DE-AC02-98CH10886 at BNL. The use of U2A and X17C beamline was supported by NSF (DMR-0805056; EAR 06-49658, COMPRES) and DOE/NNSA (DE-FC03-03N00144, CDAC).

\clearpage

\setcounter{figure}{0}
\setcounter{table}{0}
\setcounter{page}{1}
\pagenumbering{arabic}

\noindent
\textbf{Supplemental Material}

\section{1. S\lowercase{tructural phase transitions}}
\begin{figure*}[ht!]
\renewcommand{\thefigure}{S\arabic{figure}}
\includegraphics[scale=0.98]{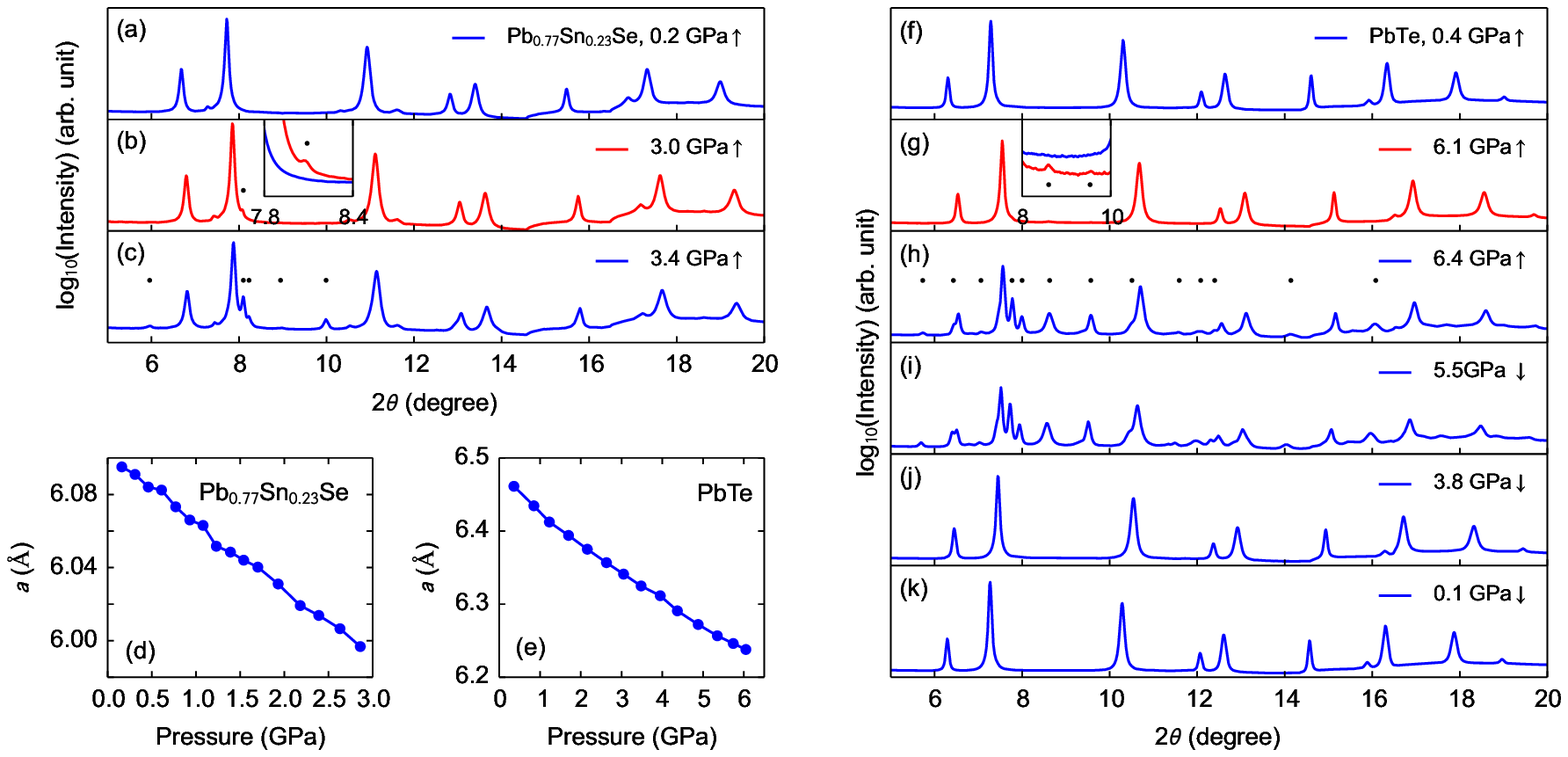}              
\caption{(a--c) XRD patterns of Pb$_{0.77}$Sn$_{0.23}$Se at room temperature. The inset in (b) compares the data at 0.2 GPa and 3.0 GPa between 7.8--8.4$^\circ$. (d,e) Pressure dependence of the lattice parameter $a$ for Pb$_{0.77}$Sn$_{0.23}$Se and PbTe. (f--k) XRD patterns of PbTe at room temperature. The inset in (g) compares the data at 0.4 GPa and 6.1 GPa between 8--10$^\circ$. The black dots indicate new peaks associated with the structural phase transitions. The arrows in the legends indicate pressure increase ($\uparrow$) or release ($\downarrow$). The incident x-ray wavelength was 0.4094 \AA.} 
\label{FIGS1}
\end{figure*}

Pb$_{0.77}$Sn$_{0.23}$Se and PbTe were investigated by powder x-ray diffraction (XRD) using a diamond anvil cell at room temperature, performed at Beamline X17C of the National Synchrotron Light Source. A 4:1 methanol-ethanol mixture was used as the pressure transmitting medium and ruby fluoresence for pressure calibration. The results are shown in Fig.~\ref{FIGS1}. A pressure-induced structural phase transition in both samples is indicated by new peaks in the powder diffraction pattern.

\begin{figure*}[htbp]
\renewcommand{\thefigure}{S\arabic{figure}}
\includegraphics[width=0.56\textwidth]{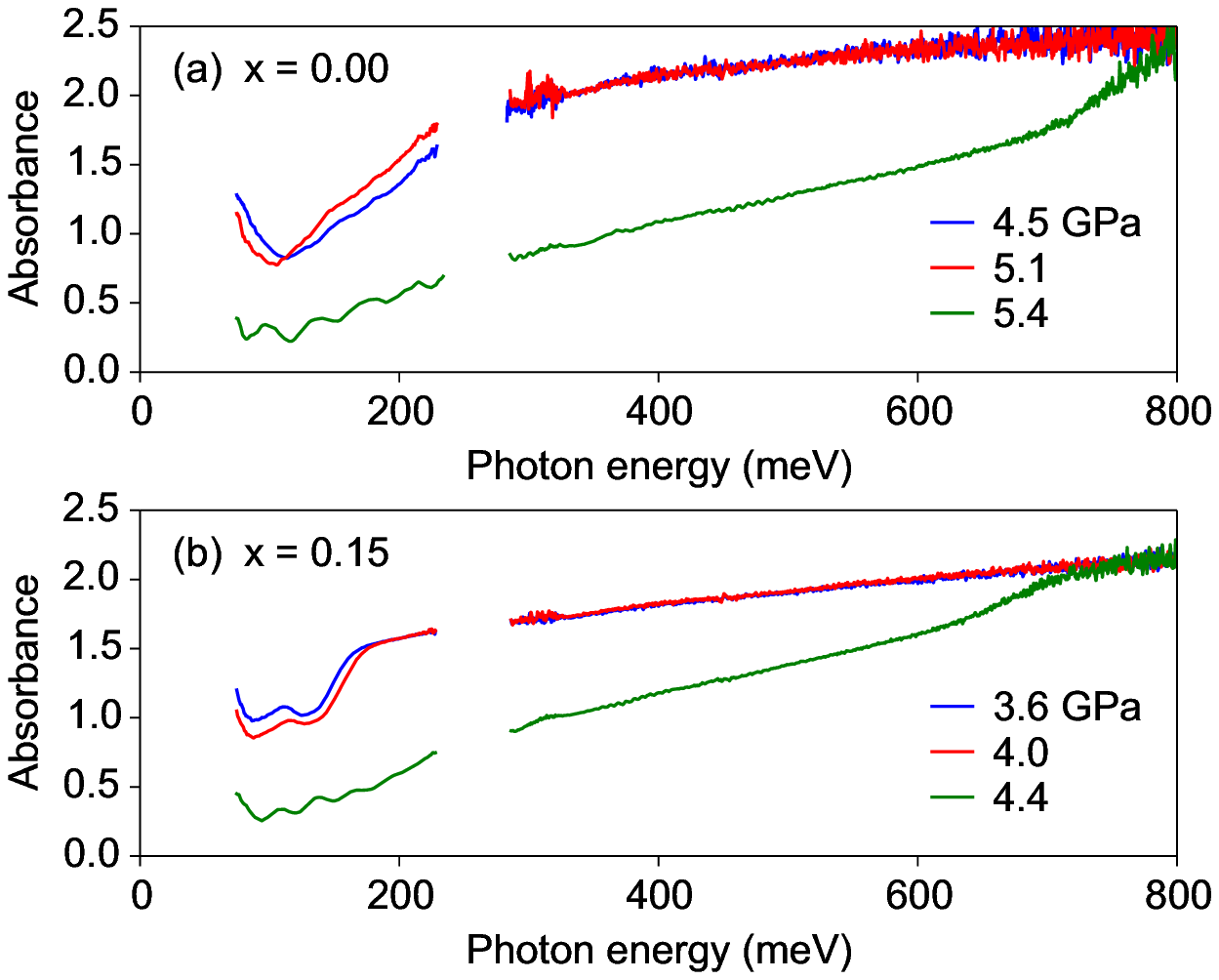}              
\caption{Infrared absorbance of (a) PbSe and (b) Pb$_{0.85}$Sn$_{0.15}$Se. A structural phase transition dramatically reduces the absorbance below 800 meV.} 
\label{FIGS2}
\end{figure*}

Concomitant with the structural phase transition are a reduction of the infrared reflectance and a dramatic decrease of the infrared absorbance, shown for PbTe in Fig.~\ref{FIGS4}(a) and (e). This was also observed in other materials, such as PbSe and Pb$_{0.85}$Sn$_{0.15}$Se; Fig.~\ref{FIGS2} shows their absorbance close to the structural phase transition. Using either XRD or infrared spectroscopy, we found that at room temperature, the ambient-pressure structure of Pb$_{1-x}$Sn$_x$Se ($x=0.00$, 0.15, and 0.23) and PbTe sustains up to 5.1, 4.0, 2.9, and 5.7 GPa, respectively.

The pressure dependence of the lattice parameter $a$ of Pb$_{0.77}$Sn$_{0.23}$Se and PbTe was obtained from Rietveld refinement on the XRD patterns, shown in Fig.~\ref{FIGS1} (d) and (e). The lattice parameter decreases almost linearly, with $a(P=0) = 6.101\pm0.001$~\AA~and $da/dP = - 0.0368\pm 0.0006$~\AA/GPa for Pb$_{0.77}$Sn$_{0.23}$Se and $a(P=0) = 6.463\pm 0.003$~\AA~and $da/dP = -  0.0386\pm0.0008$~\AA/GPa for PbTe. 

\section{2. D\lowercase{rude weight}}
The Drude weight tensor $\omega_{p,\alpha\beta}^2$ has the form \cite{Claudia2006}
\begin{equation}
\omega_{p,\alpha\beta}^2=\frac{\hbar^2e^2}{\pi m^2}\sum_n\int d\mathbf{k}\: p_{\alpha;n,n,\mathbf{k}}(\mathbf{k}) p_{\beta;n,n,\mathbf{k}}(\mathbf{k})\delta(E_n(\mathbf{k})-E_F),\label{eqs1}\tag{S1}
\end{equation}
where $\hbar$ is the Planck constant, $e$ the electron charge, $m$ the free electron mass, $n$ the band index, $\mathbf{k}$ the crystal momentum, $E_n(\mathbf{k})$ the band dispersion, and $E_F$ the Fermi energy. $p_{\alpha;n,n,\mathbf{k}}$ is the $\alpha$th component of the momentum matrix element between states in the \textit{n}th band with the crystal momentum $\mathbf{k}$,
\begin{align}
p_{\alpha;n,n,\mathbf{k}} &= \langle n,\mathbf{k}|-i\hbar \nabla_{\alpha}|n,\mathbf{k}\rangle \nonumber\\ &= \int d\mathbf{r} \: \psi_{n,\mathbf{k}}^*(\mathbf{r})(-i\hbar\nabla_{\alpha})\psi_{n,\mathbf{k}}(\mathbf{r}),\label{eqs2}\tag{S2}
\end{align}
where $\psi_{n,\mathbf{k}}(\mathbf{r})$ is the wavefunction of Bloch electrons in the $n$th band. It can be shown that \cite{Ashcroft1976}
\begin{equation}
\int d\mathbf{r} \: \psi_{n,\mathbf{k}}^*(\mathbf{r})(-i\hbar\nabla_{\alpha})\psi_{n,\mathbf{k}}(\mathbf{r}) = \frac{m}{\hbar}\frac{\partial E_{n}(\mathbf{k})}{\partial k_{\alpha}} = m v_{\alpha;n}(\mathbf{k}),\label{eqs3}\tag{S3}
\end{equation}
in which $v_{\alpha;n}(\mathbf{k})$ is the $\alpha$th component of the Bloch electron mean velocity in the $n$th band 
\begin{equation}
\mathbf{v}_n(\mathbf{k}) =\frac{1}{\hbar}\frac{\partial E_n(\mathbf{k})}{\partial \mathbf{k}}.\label{eqs4}\tag{S4}
\end{equation}
Substituting Eqs.~\eqref{eqs2} and~\eqref{eqs3} into Eq.~\eqref{eqs1}, we have
\begin{equation}
\omega_{p,\alpha
\beta}^2=\frac{\hbar^2e^2}{\pi}\sum_n\int d\mathbf{k}\: v_{\alpha;n}(\mathbf{k}) v_{\beta;n}(\mathbf{k})\delta(E_n(\mathbf{k})-E_F).\label{eqs5}\tag{S5}
\end{equation}
Evaluating the integral using the property of the Dirac delta function, the above equation can be further written as
\begin{align}
\omega_{p,\alpha\beta}^2&=\frac{\hbar^2e^2}{\pi}\sum_n (v_F)_{\alpha;n}(v_F)_{\beta;n} \int d\mathbf{k}\:\delta(E_n(\mathbf{k})-E_F)\nonumber \\
& = \frac{\hbar^2e^2}{\pi}\sum_n (v_F)_{\alpha;n}(v_F)_{\beta;n}\: g_n(E_F)\label{eqs6}\tag{S6},
\end{align}
where $(v_F)_{\alpha;n}$ is the $\alpha$th component of the Fermi velocity in the \textit{n}th band and $g_n(E_F)$ is the density of states at Fermi energy in the \textit{n}th band.

\section{3. D\lowercase{ensity of states}}
\begin{figure*}[htbp]
\renewcommand{\thefigure}{S\arabic{figure}}
\includegraphics[width=0.6\textwidth]{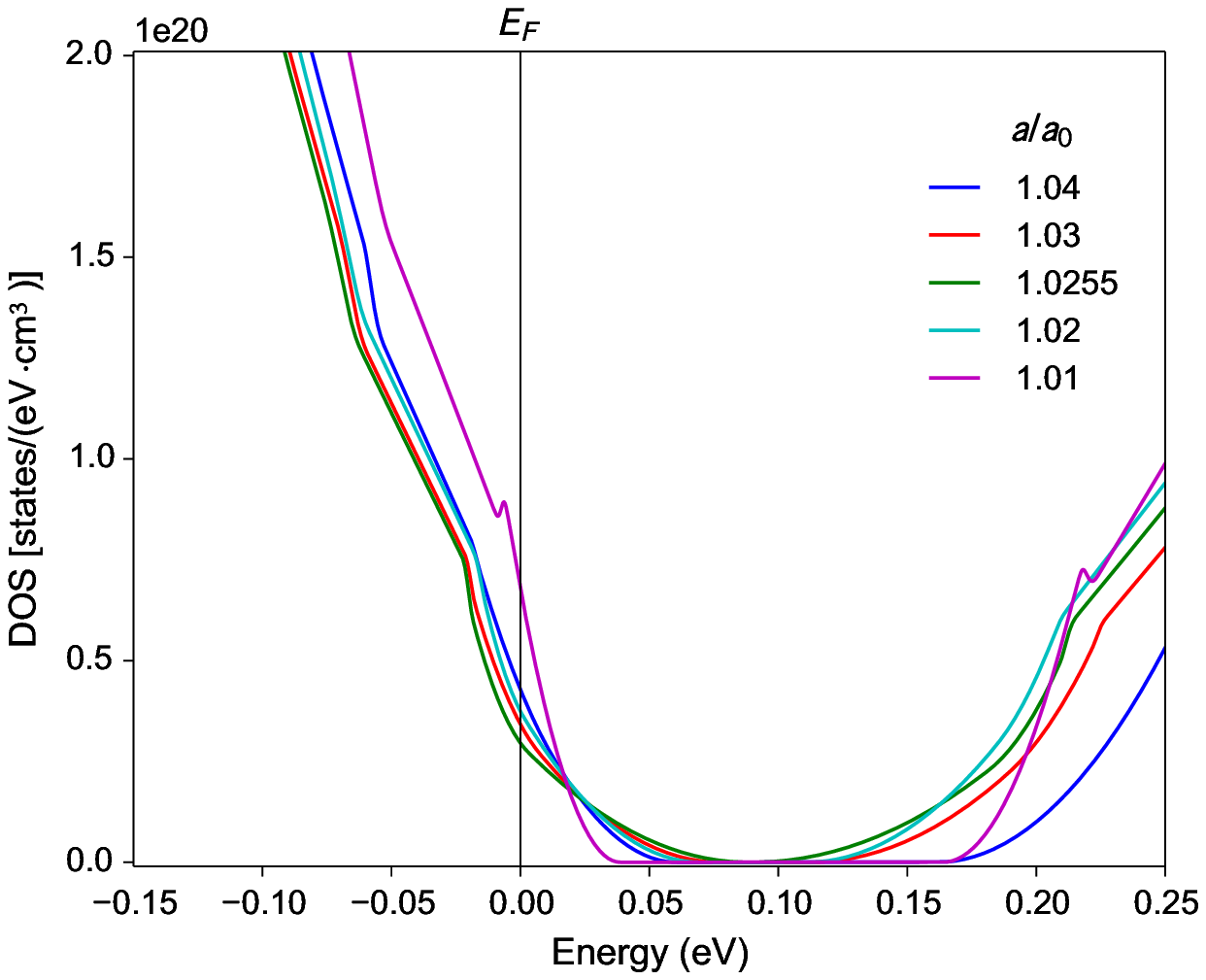}              
\caption{Electronic density of states at various $a/a_0$ corresponding to the band structure in Fig. 1(a--b) of the main text. Zero energy is defined at the Fermi level.} 
\label{FIGS6}
\end{figure*}

The electronic density of states as a function of $a/a_0$, corresponding to the band structure in Fig.~1(a--b) of the main text, is shown in Fig.~\ref{FIGS6}. Since the density of states at the Fermi energy has a minimum at the gap closing pressure, the maximum in $\omega_p$ must be attributed to the maximum in the Fermi velocity.

\section{4. F\lowercase{it parameters for reflectance}}
\begin{figure*}[htbp]
\renewcommand{\thefigure}{S\arabic{figure}}
\includegraphics[width=0.6\textwidth]{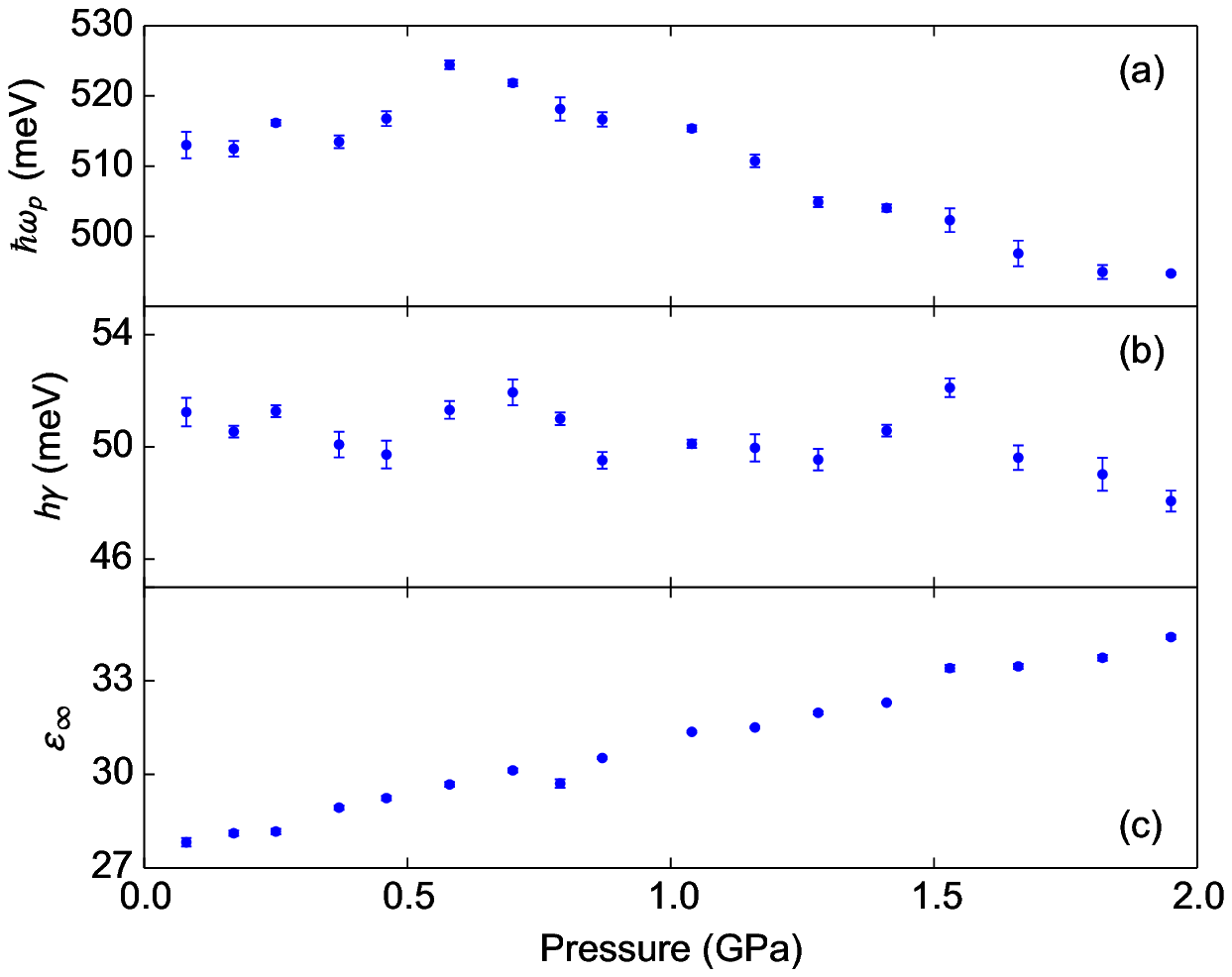}              
\caption{Fitting parameters for Pb$_{0.77}$Sn$_{0.23}$Se. (a) Plasma frequency $\omega_p$ ($\hbar\omega_p$ in meV). (b) Electronic scattering rate $\gamma$ ($h\gamma$ in meV). (c) High-energy dielectric constant $\varepsilon_{\infty}$. The error bars represent distributions of the values analyzed from three measurements.} 
\label{FIGS3}
\end{figure*}

The reflectance measured at the sample-diamond interface is approximately its bulk value because we used samples thicker than 10 $\mu$m. It is given by
\begin{equation}
R_{sd} = \frac{(n-n_d)^2+\kappa^2}{(n+n_d)^2+\kappa^2},\tag{S7}\label{Rsd}
\end{equation}
where the diamond refractive index $n_d$ is known, and the sample refractive index $n$ and extinction coefficient $\kappa$ can be obtained from the dielectric function
\begin{equation}
\varepsilon(\omega)=\varepsilon_{\infty}-\frac{\omega_p^2}{\omega^2+i\omega\gamma}.\tag{S8}\label{eps}
\end{equation}
The reflectance data of Pb$_{0.77}$Sn$_{0.23}$Se are fitted to the above equations and the fitting parameters are shown in Fig.~\ref{FIGS3}. The plasma minimum occurs roughly at the screened plasma frequency $\omega_p/\sqrt{\varepsilon_{\infty}}$, which has a maximum according to Fig.~2(c) in the main text. Since Fig.~\ref{FIGS3}(c) shows a monotonic increase of $\varepsilon_{\infty}$ with pressure, the maximum in the screened plasma frequency must be due to a maximum in $\omega_p$. This is consistent with Fig.~\ref{FIGS3}(a). 

\section{5. I\lowercase{nfrared data for} P\lowercase{b}T\lowercase{e}}
\begin{figure*}[ht]
\renewcommand{\thefigure}{S\arabic{figure}}
\includegraphics[width=0.98\textwidth]{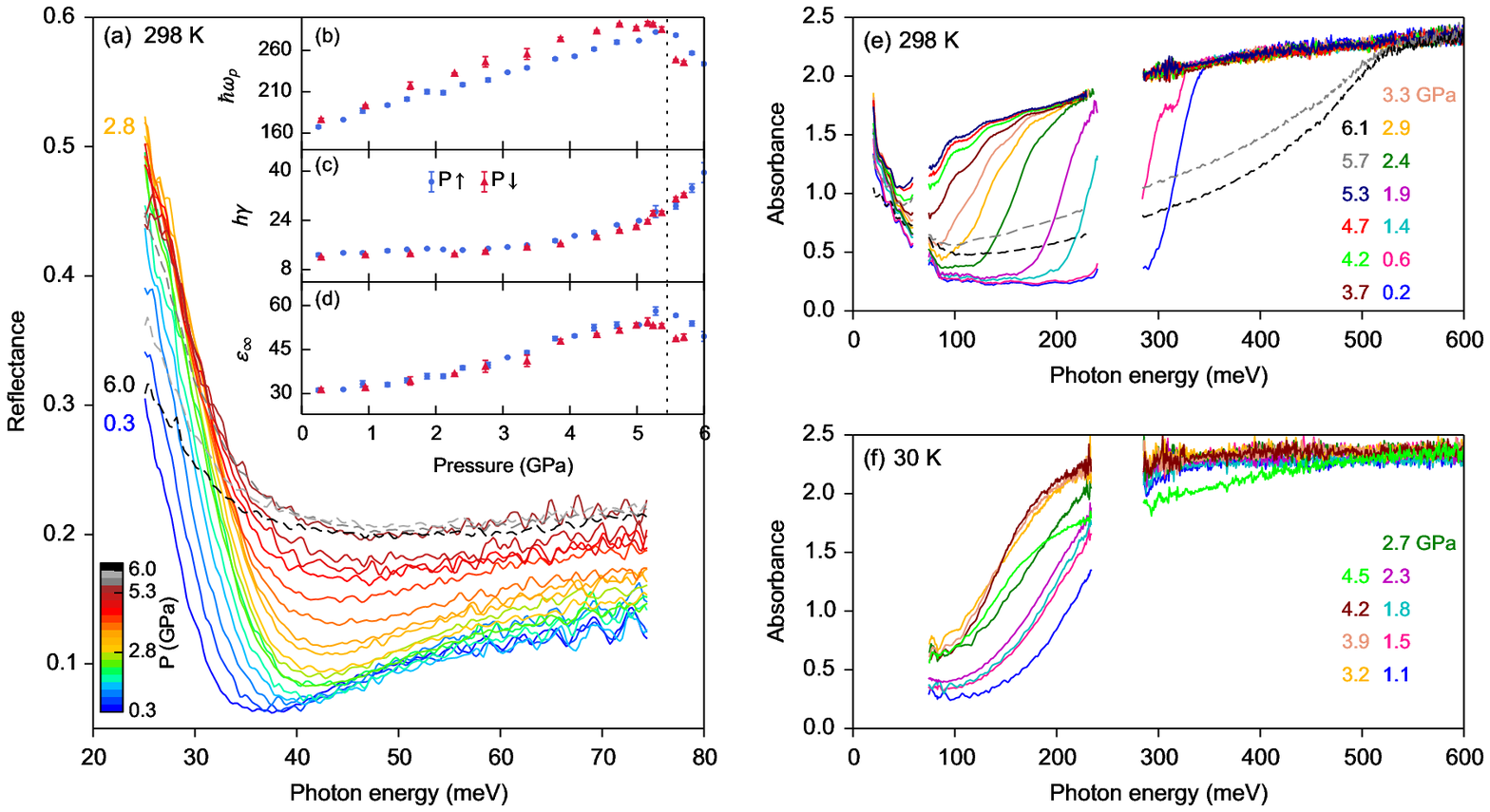}              
\caption{PbTe under pressure. (a) Far-infrared reflectance at room temperature. The dashed lines show the reflectance after the structural phase transition. The insets are pressure dependence of the (b) plasma frequency $\omega_p$ ($\hbar\omega_p$ in meV), (c) electronic scattering rate $\gamma$ ($h\gamma$ in meV), and (d) high-energy dielectric constant $\varepsilon_{\infty}$. Pressure was increased to 6 GPa and then released. Results for pressure increase (P$\uparrow$) and pressure release (P$\downarrow$) are compared. The error bars represent distributions of the values analyzed from three measurements. The dotted line indicates the structural phase transition. (e) Far-infrared and mid-infrared absorbance measured at room temperature on two samples. The far-infrared data were measured using Vaseline as the pressure-transmitting medium. (f) Mid-infrared absorbance measured at 30 K. The blank region in (e) and (f) between 240--280 meV corresponds to strong diamond absorption.} 
\label{FIGS4}
\end{figure*}

We also performed high-pressure infrared reflectance and absorbance measurements on PbTe. The results are shown in Fig.~\ref{FIGS4}. The far-infrared reflectance data are fitted to Eqs. (S7) and (S8). $\varepsilon_{\infty}$ [Fig.~\ref{FIGS4}(d)] increases monotonically but drops at about 5.6 GPa due to the structural phase transition. The second structure persists even when the pressure is released to below 5.6 GPa, resulting in a hysteresis loop in $\varepsilon_{\infty}$. Although $\omega_p$ [Fig.~\ref{FIGS4}(b)] has a maximum at about 5.3 GPa, it is too close to the structural phase transition to conclude whether a topological phase transition is the origin. The absorption edge in the room-temperature infrared absorbance [Fig.~\ref{FIGS4}(e)] systematically redshifts up to 5.3 GPa and then dramatic changes its shape at 5.7 GPa, associated with the structural phase transition. In the same pressure range, the absorption peak in the far infared grows and saturates, followed by a precipitous drop due to the structural phase transition. 

To separate the intraband and interband transitions, we cooled the sample to 30 K and measured the mid-infrared absorbance, shown in Fig.~\ref{FIGS4}(f). The absorption edge first redshifts from 1.1 to 3.2 GPa; it then tilts towards higher photon energy at 3.9 and 4.2 GPa, followed by a structural phase transition at 4.5 GPa. This is similar to that observed in PbSe, serving as another example of pressure-induced band gap closing and reopening.

\section{6. A\lowercase{dditional band structure data}}
\begin{figure*}[htbp]
\renewcommand{\thefigure}{S\arabic{figure}}
\includegraphics[width=0.58\textwidth]{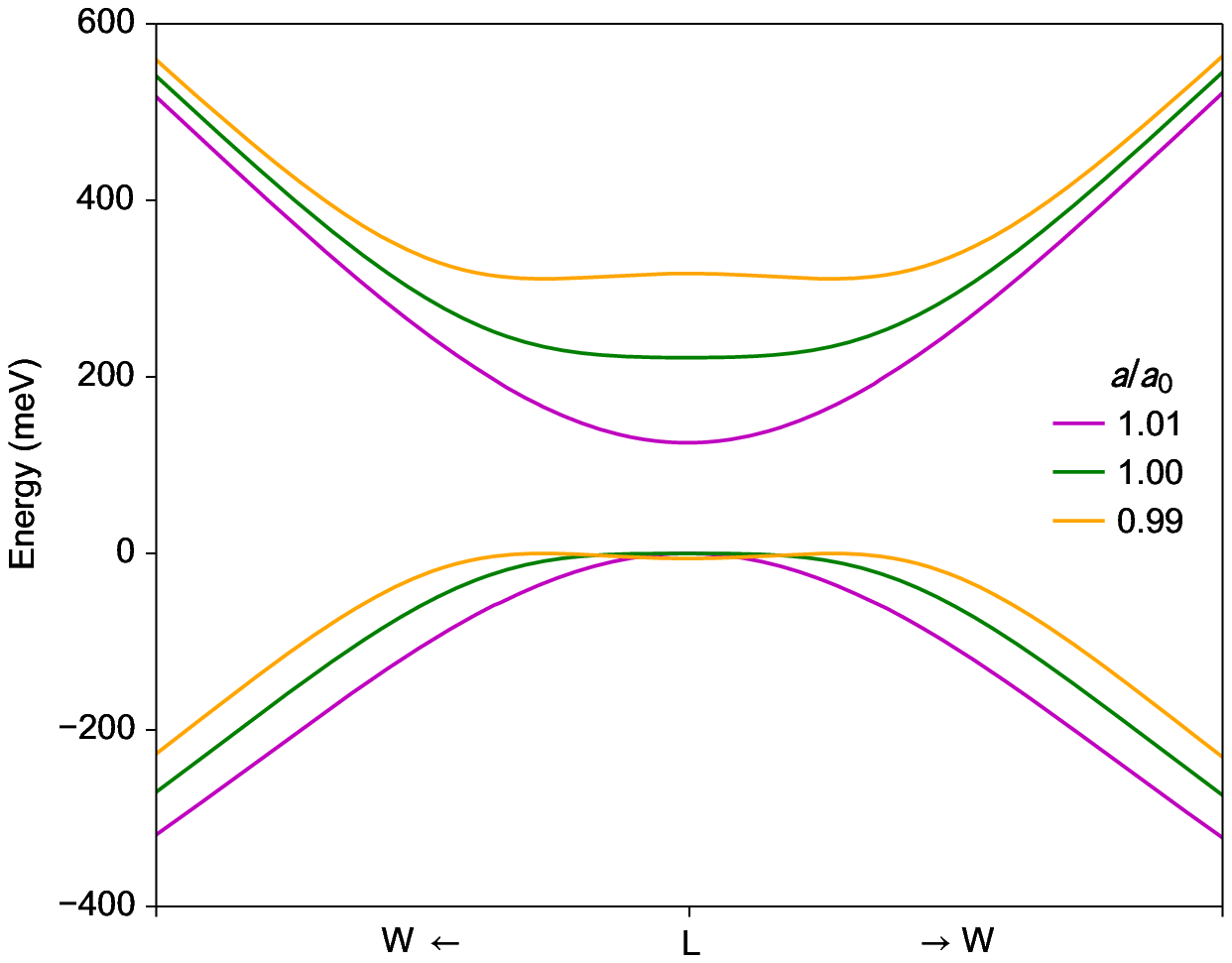}              
\caption{Electronic band structure of PbSe along the L-W direction of the Brillouin zone for lattice parameter ratio $a/a_0=1.01$, 1.00 and 0.99, where $a_0$ is the experimental lattice constant in the zero-temperature limit and at ambient pressure.} 
\label{FIGS5}
\end{figure*}

The calculated electronic band structure of PbSe near the L point of the Brillouin zone shows a Mexican hat shape at sufficiently high pressure, as shown in Fig.~\ref{FIGS5} at $a/a_0=0.99$.

\begin{thebibliography}{99}
\bibitem{fringe} Raw data in Fig.~\ref{FIG2}(a) below 2.0 GPa and in Fig.~\ref{FIG3}(a--b) contain interference fringes due to the multiple reflections of the infrared beam in the sample and pressure-transmitting medium. For clarity, these fringes are removed by Fourier transforming the spectra and filtering the associated peaks.
\bibitem{Note} The spectral features for absorbance greater than 2.2 are likely artifacts due to stray light leakage, but they do not affect the analysis discussed here.
\end{thebibliography}
\end{document}